\documentclass[twocolumn,
aps,nofootinbib,showpacs,showkeys,preprint
tightenlines
] {revtex4}
\usepackage{epsf,epsfig,subfigure,graphicx,amsmath,amssymb}
\usepackage{color}
\usepackage{float}
\newcommand{\dis}[1]{\begin{equation}\begin{split}#1\end{split}\end{equation}}

\newcommand{\ie}{{\it i.e.}\ }
\newcommand{\etal}{{\it et al.}}

\newcommand{\Od}{{\cal{O}}}
\newcommand{\Vp}{V_{\rm PMNS}}

\begin{document}

\title{\large\bf  Parametrization of PMNS matrix based on dodeca-symmetry}

\author{Jihn E. Kim$^{a,b}$ and Min-Seok Seo$^a$
\email{jekim@ctp.snu.ac.kr}
}
\affiliation{
$^a$Department of Physics and Astronomy and Center for Theoretical Physics, Seoul National University, Seoul 151-747, Korea\\
$^b$GIST College, Gwangju Institute of Science and Technology,  Gwangju 500-712, Korea \\
 }
\begin{abstract}
The dodeca symmetry is designed to obtain the Cabibbo angle $\theta_C^{\rm CKM}$ approximately $15^{\rm o}$ and the (11) element of $\Vp$ as $\cos  30^{\rm o}$, leading to $\theta_1^{\rm PMNS}+\theta_C^{\rm CKM}\simeq 45^{\rm o}$. This leading order dodeca symmetric $\Vp$ is corrected by small parameters, especially as an expansion in terms of a small parameter $\beta$. Neglecting two Majorana phases, the expression of $\Vp$ contains four parameters: a small $\beta$, and three $\Od(1)$ parameters $A,B,$ and $\delta$. From the neutrino oscillation data, we present two parametrizations and estimate their $\beta$'s.
\end{abstract}


\maketitle

\section{Introduction}\label{sec:Introduction}

Nonzero neutrino masses and oscillation observed in the last two decades have led to a large mixing in the leptonic sector \cite{NeutrinoReview}, unlike in the quark sector. The leptonic mixing known as the Pontecorvo-Maki-Nakagawa-Sakata (PMNS) matrix \cite{PMNS} takes a quite different form from that of Cabbibo, Kobayashi and Maskawa(CKM) \cite{CKM73}. The PMNS matrix takes a bi-maximal mixing, whereas the CKM matrix is close to the identity. So, the CKM matrix being close to the identity has the famous Wolfenstein approximate form \cite{Wol83}. In this paper, we investigate two approximate forms of the PMNS matrix, following the method of the quark mixing presented recently in Ref. \cite{KimSeo11} where the CP violation is directly noticeable from the unitary matrix itself.

The PMNS matrix, being not close to identity, triggered a tremendous theoretical activity to obtain a leading bi-maximal form. Most studies are based on non-Abelian discrete symmetries started with the permutation symmetry $S_3$ since the late 1970s \cite{Perm70s}.

The most recent popular leptonic mixing pattern is the so-called
tri-bimaximal mixing, suggested by Harrison, Perkins, and Scott in
2002 \cite{tribi}, which in addition to the bi-maximal mixing in the
(23) and(33) elements the another column for the solar mixing angle
is given by $\theta_{\mathrm{sol}}=\sin^{-1}\frac{1}{\sqrt3}$, with the vanishing (13) element. This pattern can be
derived by using various permutation symmetries and their subgroups
such as $S_4$ or $A_4$. At the face value, the tri-bimaximal mixing
looks like being very close to the current best fit(BF) values
\cite{Albright10}. However, the very closeness to the BF values
might be a dilemma in that there is not much room for an adjustable
parameter if more accurate data produce a significant deviation from
the tri-bimaximal pattern. In this sense, since the recently proposed
dodeca form \cite{KimSeo10}, which produces a bi-maximal
but is not as close to the BF values as the tri-bimaximal form
\cite{Albright10}, has a room to introduce a relatively large
correction parameter $\beta$.\footnote{We call the expansion
parameter $\beta$ (meaning around the {\it bi}-maximal),
corresponding to the expansion parameter $\lambda$ (around the
identity) of Wolfenstein \cite{Wol83}.} The dodeca pattern is
$N=12$ case of the dihedral groups $D_N$ studied for neutrino masses
in \cite{Blum08,Review10}. One nice feature of the dodeca pattern is
that it gives $\theta_{\rm sol}^{\rm PMNS}=\frac{\pi}{6}$ and
$\theta_C^{\rm CKM}=\frac{\pi}{12}$ \cite{KimSeo10}, producing a
cherished relation, $\theta_{\rm sol}^{\rm PMNS}+\theta_C^{\rm
CKM}\simeq \frac{\pi}{4}$ \cite{Park06}.

The fact that the PMNS matrix takes a very different form from the
identity implies that we have to parametrize it in a different way
from what Wolfenstein did for the CKM matrix, the expansion around
the identity. In fact, Qin and Ma tried the expansion of the PMNS
matrix around $\frac{\pi}{4}$ \cite{QinMaPRD11}.
Taking the relation  $\theta_{\rm sol}^{\rm PMNS}+\theta_C^{\rm
CKM}\simeq \frac{\pi}{4}$ into account, they expanded the PMNS matrix in terms of  $\lambda$, the expansion parameter of the CKM matrix, around the tri-bimaximal mixing point \cite{QinMaPRD11,AhnCh11}.
It is an attractive
idea to expand the small parameters around the basic pattern
originating from a non-Abelian discrete symmetry. This expansion
parameter is better not to be too small as commented above. In this
sense, we expand the PMNS matrix around the dodeca
pattern.

Since the deviation of (23) and (33) element from
bi-maximality is almost negligible \cite{Valle08}, \ie statistically
insignificant, we require that the deviation of the atmospheric mixing angle (the
$\nu_\mu-\nu_\tau$ mixing) from $\frac{\pi}{4}$ is very small.
The (13) element is known to be very small, and is zero in the tri-bimaximal and dodeca mixing. On the other hand, from the solar neutrino data, a maximal solar neutrino mixing, \ie
$\theta_{12}=\frac{\pi}{4}$, is ruled out at more than 6$\sigma$
\cite{PData10}, and one has $\cos 2 \theta_{12} \geq 0.26$ (at
99.73$\,\%$ C.L.). Therefore, the solar neutrino angle deviation
from $\frac{\pi}{4}$ or $\frac{\pi}{6}$ (the case of dodeca mixing)
can taken as an expansion parameter $\beta$.

Most weak CP violating processes of the light leptons, except in the neutrinoless double beta decay \cite{Valle82double}, do not depend on two Majorana phases of the PMNS matrix. Therefore, let us focus on the PMNS matrix, neglecting two Majorana phases. Then, the nonzero (13) elements gives a barometer of the CP violation in the leptonic sector as formulated and emphasized in Ref. \cite{KimSeo11}. Our expansion in terms of $\beta$ is given in
two plausible cases for the (13) element, one proportional to $\beta^2$ and the
other proportional to $\beta$.  Since the (31) and (22) elements are
of order $\Od(1)$, the Jarlskog determinant \cite{Jarlskog85}
following the method of Ref. \cite{KimSeo11}  should be in the same
order as the (13) element, $\beta^2$ or $\beta$ depending our
models. Therefore, looking for the deviation of (13) element from
zero is an important barometer of the leptonic CP violation, which
triggered the worldwide experimental activities to measure the (13)
element of $\Vp$ \cite{RENO}.

\section{measurements of the PMNS matrix elements}\label{sec:exp}

Unlike the CKM case, the PMNS matrix elements have not been determined very accurately.
To set up the discussion of the next section, we briefly present the determination of the PMNS matrix element\cite{PData10}. The Chau-Keung-Maiani type parametrization is
     \begin{widetext}
     \dis{ \left(
\begin{array}{ccc}
  c_{12}c_{13},& s_{12}c_{13},&s_{13}e^{-i \delta} ,\\[1 em]
  -s_{12}c_{23}-c_{12}s_{23}s_{13}e^{i \delta},&
  c_{12}c_{23}-s_{12}s_{23}s_{13}e^{i \delta}   &s_{23}c_{13}
 \\ [1em]
 s_{12}s_{23}-c_{12}c_{23}s_{13}e^{i \delta} ,& -c_{12}s_{23}-s_{12}c_{23}s_{13}e^{i \delta}     ,
  & c_{23}c_{13}
  \end{array}\right)\label{eq:CKM}
  }
     \end{widetext}
where $c_{ij}=\cos\theta_{ij}$ and $s_{ij}=\sin\theta_{ij}$. The four parameters are
$\theta_{12}, \theta_{23}, \theta_{13}$, and the phase $\delta$. To relate directly to the neutrino oscillation experiments, we also use the notation $\theta_{{\rm atm}}$ instead of $\theta_{23}$ and $\theta_{{\rm sol}}$ instead of $\theta_{12}$.
The flavor basis $\vert \nu_\alpha \rangle$ is the superposition of the mass eigenstates $\vert \nu_j \rangle$ with the coefficients provided by the PMNS matrix elements,
\dis{\vert \nu_\alpha \rangle= \sum_{j}V^*_{\alpha j} \vert \nu_j ;p_j\rangle
}
where the subscript $\alpha$ indicates $e, \mu, \tau$, and $j$ runs from 1 to 3. It is the convention that take $\nu_1$ as the lightest neutrino. The probability amplitude of observing $\vert \nu_{\alpha^{\prime}} \rangle$ after the propagation in spacetime interval $(T,L)$ of $\vert \nu_\alpha \rangle$ is given by
\dis{A(\nu_\alpha \to \nu_{\alpha^{\prime}})=\sum_{j} V_{\alpha^{\prime}j}V_{j\alpha}^{\dagger} e^{-i(E_j T -p_j L)} }
and the probability $P(\nu_\alpha \to \nu_{\alpha^{\prime}})$ is just $\vert A(\nu_\alpha \to \nu_{\alpha^{\prime}}) \vert^2$,
\dis{P&(\nu_\alpha \to \nu_{\alpha^{\prime}}) \cong P(\bar{\nu}_\alpha \to \bar{\nu}_{\alpha^{\prime}})
   \\& \cong \delta_{\alpha\alpha^{\prime}}-2 \vert V_{l3}\vert^2(\delta_{\alpha\alpha^{\prime}}-\vert V_{\alpha^{\prime}3}\vert^2)(1-\cos\frac{\Delta m^2_{31}}{2p}L).}
Therefore, the suvival of the electron type neutrino is given by
\dis{P(\nu_e \to \nu_e) \cong \delta_{\alpha\alpha^{\prime}}-2 \vert V_{e3}\vert^2(1-\vert V_{e3}\vert^2)(1-\cos\frac{\Delta m^2_{31}}{2p}L)}
which is used in the CHOOZ, Daya Bay and RENO experiments. A similar expression can be written for $P(\nu_{\mu} \to \nu_{\mu})$, used in K2K and MINOS experiments.
On the other hand, the appearance is given by
\dis{P(\nu_{\mu (e)} \to \nu_{e(\mu)})& \cong 2  \vert V_{\mu3}\vert^2\vert V_{e3}\vert^2(1-\cos\frac{\Delta m^2_{31}}{2p}L)
   \\
   & =\frac{\vert V_{\mu 3}\vert^2}{1-\vert V_{e 3}\vert^2} P^{2 \nu}(\vert V_{e 3}\vert^2, \Delta m^2_{31})}
where $P^{2 \nu}(\vert V_{e 3}\vert^2, \Delta m^2_{31})$ indicates the probability of 2-neutrino transition, $\nu_e \to (s_{{\rm atm}}\nu_{\mu}+c_{{\rm atm}}\nu_{\tau})$, used in MINOS experiment. Similar expression for $P(\nu_{\mu} \to \nu_{\tau})$ is used in OPERA.

When the neutrino source has a sizable dimension $\Delta L$ and the energy resolution of detector is $\Delta E$, we integrate over the region of neutrino source and energy resolution function. Then, a large phase $\frac{\Delta m^2_{31}}{2p}L$ in the argument of $\cos$ is averaged over and the average probability is given by
\dis{\overline{P}(\nu_\alpha \to \nu_{\alpha^{\prime}})=\overline{P}(\bar{\nu}_\alpha \to \bar{\nu}_{\alpha^{\prime}})\cong \sum_{j} \vert V_{\alpha^{\prime}j} \vert^2 \vert V_{\alpha j} \vert^2.}
Especially, for the case of $\alpha=\alpha^{\prime}=e$, the averaged probability is
\dis{\overline{P}(\nu_e \to \nu_e)&=\overline{P}(\bar{\nu}_e \to \bar{\nu}_e)
    \\
    &\cong \vert V_{e3} \vert^4 + (1-\vert V_{e3} \vert^2)^2 P^{2 \nu}(\nu_e \to \nu_e) }
where
\dis{P^{2 \nu}(\nu_e \to \nu_e)&=P^{2 \nu}(\bar{\nu}_e \to \bar{\nu}_e)
\\
& =1-\frac{1}{2}\sin^2 2 \theta_{{\rm sol}}(1-\cos\frac{\Delta m^2_{21}}{2p}L),}
which has been used in the KamLand experiments.

The solar neutrino angle $\theta_{{\rm sol}}$ can be determined from the solar neutrino flux observation, for example, in the SNO and the Super-Kamiokande experiments or from the detection of $\bar{\nu}_e$ neutrinos emitted from the nuclear power reactors in the KamLand. The atmospheric neutrino angle $\theta_{{\rm atm}}$ measurement can be made by observing the atmospheric neutrino, the product of cosmic ray interaction in the atmosphere, in Super-Kamiokande, or product from accelerator experiment, for example, in the K2K and MINOS experiments. Finally, the deviation from zero of $V_{13}$ is determined by observing $P(\bar{\nu}_e \to \bar{\nu}_e)$ in the CHOOZ experiment, and $P(\nu_{\mu} \to \nu_e)$ in the K2K experiment.

The leptonic CP violation is significant for the following asymmetry in the neutrino oscillation \cite{PData10},
\dis{A_{CP}^{\alpha^{\prime} \alpha} & \equiv P(\nu_\alpha \to \nu_{\alpha^{\prime}})-P(\bar{\nu}_\alpha \to \bar{\nu}_{\alpha^{\prime}})
\\
& =4\sum_{j>k}{\rm Im}(V_{\alpha^{\prime}j} V^*_{\alpha j} V_{\alpha k} V^*_{\alpha^{\prime}k} )
\sin \frac{\Delta m^2_{jk}}{2p}L  }
Especially,
\dis{ A_{CP}^{(\mu e)}&=-A_{CP}^{(\tau e)}=A_{CP}^{(\tau \mu )}
\\
&=4J (\sin\frac{\Delta m^2_{32}}{2p}L+\sin\frac{\Delta m^2_{21}}{2p}L+\sin\frac{\Delta m^2_{13}}{2p}L) }
where $J={\rm Im}(V_{\alpha^{\prime}j} V^*_{\alpha j} V_{\alpha k} V^*_{\alpha^{\prime}k} )$ is the Jarlskog determinant \cite{Jarlskog85} of the  PMNS matrix up to the sign.
Note that if all three neutrinos are degenerate, there is no CP violation. It is because one can find a unitary matrix to make the PMNS matrix identity for the degenerate neutrino masses.

With the above experiments,  the best fit values(BF) and 99.73$\%$ C.L. values are listed as follows\cite{PData10}:
\dis{
 &7.05\times 10^{-5}{\rm eV}^2\leq\Delta m_{12}^2\leq 8.34\times 10^{-5}{\rm eV}^2\\
  &0.25 \leq \sin^2 \theta_{12} \leq 0.37\\
   &2.70\times 10^{-3}{\rm eV}^2\leq \vert\Delta m_{31}^2 \vert\leq 2.75\times 10^{-3}{\rm eV}^2\\
  &0.36 \leq \sin^2 \theta_{23} \leq 0.67\\
  &\sin^2 \theta_{13}<0.035(0.056)~~~{\rm at~90\%~(99.73\%)~C.L.}
  }
with the following BF values
\dis{
&(\Delta m_{12}^2)_{BF}=7.65\times 10^{-5}{\rm eV}^2,\\
&(\sin^2\theta_{12})_{BF}=0.304,\\
&(\vert \Delta m_{31}^2 \vert)_{BF}=2.40 \times 10^{-3}{\rm eV}^2,\\
 &(\sin^2\theta_{23})_{BF}=0.5,
 }
Reference \cite{Valle08} presented the BF value of
  $\sin^2 \theta_{13}$ as 0.01 which is at $0.9\sigma$ away from $\theta_{13}=0$. At present, there is no decisive evidence for a nonzero $\theta_{13}$, but future experiments \cite{RENO} are expected to give better limits on  $\sin \theta_{13}$.
   Recently, T2K collaboration reported a large $\theta_{13}$ \cite{T2K}. At the 90\% confidence limit, they report $0.03(0.04)<\sin^2 2\theta_{13}<0.28(0.34)$ for $\sin^2 2\theta_{23}=1.0,~\vert \Delta m^2_{23} \vert =2.4 \times 10^{-3} {\rm eV}^2,~ \delta=0$ and normal(inverted) hierarchy. The BF points are 0.11(0.14).

\section{The PMNS matrix with (13) element of $\Od(\beta^2)$}\label{sec:betaSquared}

From the discussion of Sec. \ref{sec:exp}, the quite accurate bi-maximality is an established fact. However, the tri-bimaximality or dodeca symmetry is approximate. We may start from
the exact mixing matrix adopted in \cite{KimSeo11},
\dis{ \left(
\begin{array}{ccc}
  c_1,& s_1c_3,
&s_1s_3,\\[1 em]
  -c_2s_1,&
  e^{-i \delta}s_2s_3+c_1c_2c_3   &-e^{-i \delta}s_2c_3+c_1c_2s_3
 \\ [1em]
 -e^{i \delta}s_1s_2  ,&  -c_2s_3+c_1s_2c_3 e^{i \delta}     ,
  & c_2c_3+c_1s_2s_3 e^{i \delta}
  \end{array}\right)\label{eq:KSform}
 }

The deviations from these hypothetical symmetries contain large correction factors denoted as $\beta$. Therefore, we satisfy the bi-maximality of $\nu_\mu-\nu_\tau$ mixing up to $\Od(\beta^2)$. In this section, we assume that the deviation of the (13) element from zero is rather small, \ie $(B/2)\beta^2$.
Then, we can expand $\Vp$ around the dodeca symmetric point as

\begin{widetext}

\dis{ \left(
\begin{array}{ccc}
  \frac{1}{2}(\sqrt3-\beta-\frac{\sqrt3}{2}\beta^2),& \frac{1}{2}(1+\sqrt3\beta-\frac{\beta^2}{2}),
&\frac{B}{2} \beta^2 ,\\[1 em]
  -\frac{1}{2\sqrt2}(1+\sqrt3\beta-(A+\frac{1}{2})\beta^2),&
  \frac{1}{2\sqrt2}(\sqrt3-\beta-(\sqrt3A+\frac{\sqrt3}{2}-e^{-i\delta}2B)\beta^2)   &-\frac{e^{-i\delta}}{\sqrt2}(1+(A-e^{i\delta}\frac{\sqrt3}{2}B)\beta^2)
 \\ [1em]
 -\frac{e^{i\delta}}{2\sqrt2}(1+\sqrt3\beta+(A-\frac{1}{2})\beta^2) ,& \frac{e^{i\delta}}{2\sqrt2}(\sqrt3-\beta+(\sqrt3A-\frac{\sqrt3}{2}-e^{-i\delta}2B)\beta^2)     ,
  & \frac{1}{\sqrt2}(1-(A-e^{i\delta}\frac{\sqrt3}{2}B)\beta^2)
  \end{array}\right)\label{eq:Vpbsquare}
}
The approximate PMNS matrix (\ref{eq:Vpbsquare}) satisfies the unitarity up to order $\beta^2$. As formulated in Ref. \cite{KimSeo11}, Eq.  (\ref{eq:Vpbsquare}) has the real determinant, which is required from the SU(2)$_W$ Peccei-Quinn symmetry \cite{PQ77}. Namely, we must work in a well-defined lepton basis \cite{KimSeo11}.  Then, each term of the determinant shows whether the CP violation is present or not. In particular, the product of (31), (22), and (13) elements is the barometer of CP violation \cite{KimSeo11}. The determinant of Eq.  (\ref{eq:Vpbsquare}) has the following six terms,
\dis{
V_{11}V_{22}V_{33} =&\frac{1}{8}(3-2\sqrt3\beta-(6A+2-\frac{7\sqrt3}{2}B\cos\delta+i\frac{\sqrt3}{2}B\sin \delta)\beta^2)\\
-V_{11}V_{23}V_{32} &=\frac{1}{8}(3-2\sqrt3\beta+(6A-2-\frac{7\sqrt3}{2}B\cos\delta+i\frac{\sqrt3}{2}B\sin \delta)\beta^2)\\
V_{12}V_{23}V_{31} &=\frac{1}{8}(1+2\sqrt3\beta+(2A+2-\frac{\sqrt3}{2}B(\cos\delta+i\sin \delta)\beta^2)\\
-V_{12}V_{21}V_{33} &=\frac{1}{8}(1+2\sqrt3\beta-(2A-2-\frac{\sqrt3}{2}B(\cos\delta+i\sin \delta)\beta^2)\\
V_{13}V_{21}V_{32} &=-\frac{\sqrt3}{16}B\beta^2(\cos\delta+i\sin\delta)\\
-V_{13}V_{22}V_{31} &=\frac{\sqrt3}{16}B\beta^2(\cos\delta+i\sin\delta)\,. \label{eq:elementDet}
}
\end{widetext}
We note that each term of Eq. (\ref{eq:elementDet}) has the same magnitude for the imaginary part,
\dis{
\left|\frac{\sqrt3}{16}B\beta^2  \sin\delta\right|
}
as noted in Ref. \cite{KimSeo11} for the CKM matrix. As in the CKM case,
the Jarlskog determinant of PMNS matrix can be read directly from the unitarity matrix, Eq. (\ref{eq:Vpbsquare}). The area of the Jarlskog triangle is of order the small parameter squared ($\beta^2$) and we can say that the CP violation effect in the lepton sector is larger than that in the quark sector. Note that in the quark sector the area of the Jarlskog triangle is of order the sixth power of the small parameter ($\lambda^6$) \cite{KimSeo11}.

Note that if the (13) element vanishes, all the phases in
PMNS matrix except two Majorana phases can be removed away. Assuming that PMNS matrix elements are given by their best-fit values, and $\sin \theta_{13}\cong 0.1$,  the parameters are given by
\dis{&\beta=0.068,~~~B=39,\\
& 2 A \beta^2 \cong 0.311 \cos \delta\label{eq:ABlarge}
}
When there is no CP violation($\delta=0$), $A=33$ and with maximal CP violation($\delta=\frac{\pi}{2}$),  $A=0$.
Since both $A$ and $B$ of order 10 from Eq. (\ref{eq:ABlarge}), it is better to have a parametrization for the (13) element being of order of $\beta$.
So, let us consider the parametrization
\begin{widetext}

\dis{ \left(
\begin{array}{ccc}
  \frac{1}{2}(\sqrt3-\beta-\frac{\sqrt3}{2}\beta^2),& \frac{1}{2}(1+\sqrt3\beta-\frac{\beta^2}{2}),
&\frac{B}{2} \beta^2 ,\\[1.5 em]
   \begin{array}{c}
 -\frac{1}{2\sqrt2}\Big(1+(\sqrt3-A)\beta\\
 -\frac{1}{2}(1+2\sqrt3A+A^2)\beta^2\Big)
 \end{array},&
  \begin{array}{c}
  \frac{1}{2\sqrt2}\Big(\sqrt3-(1+\sqrt3A)\beta\\
  -(\frac{\sqrt3}{2}(1+A^2)-A-e^{-i\delta}2B)\beta^2\Big)
  \end{array},
   &-\frac{e^{-i\delta}}{\sqrt2}\Big(1+A\beta-(\frac{A^2}{2}+e^{i\delta}
   \frac{\sqrt3}{2}B)\beta^2\Big)
 \\ [2.5em]
   \begin{array}{c}
 -\frac{e^{i\delta}}{2\sqrt2}\Big(1+(\sqrt3+A)\beta\\
 +(\sqrt3A-\frac{A^2}{2}-\frac{1}{2})\beta^2\Big)
 \end{array},
 &
   \begin{array}{c}
\frac{e^{i\delta}}{2\sqrt2}\Big(\sqrt3+(-1+\sqrt3A)\beta\\
-(\frac{\sqrt3}{2}(1+A^2)+A+e^{-i\delta}2B)\beta^2\Big)
\end{array} ,
  & \frac{1}{\sqrt2}\Big(1-A\beta+(-\frac{A^2}{2}+\frac{\sqrt3}{2}Be^{i\delta})\beta^2\Big)
  \end{array}\right)
}
\end{widetext}
and $A\cong2.2$ .

\section{The PMNS matrix with (13) element of $\Od(\beta)$}\label{sec:beta}

 Since the (13) element of the PMNS matrix is not measured accurately, we can consider the correction of the (13) element of $\Od(\beta)$.

As in Sec. \ref{sec:betaSquared}, we keep the deviation of $\theta_2$ from
$\pi/4$ at $\Od(\beta^2)$, and obtain the following
parametrization,

\begin{widetext}
 \dis{ \left(
\begin{array}{ccc}
  \frac{1}{2}(\sqrt3-\beta-\frac{\sqrt3}{2}\beta^2),& \frac{1}{2}\Big(1+\sqrt3\beta-\frac{1}{2}(1+B^2)\beta^2\Big),
& \frac{B}{2} (1+\sqrt3 \beta) \beta ,\\[1 em]
  -\frac{1}{2\sqrt2}\Big(1+\sqrt3\beta-[A+\frac{1}{2}]\beta^2\Big),&
  \begin{array}{c}
  \frac{1}{2\sqrt2}\Big(\sqrt3-[1-2Be^{-i\delta}]\beta\\[0.2em]
  -\frac{\sqrt3}{2}(1+2A+B^2)\beta^2\Big)
  \end{array}, &
  \begin{array}{c}
   -\frac{e^{-i\delta}}{\sqrt2}\Big(1-\frac{\sqrt3}{2}e^{i \delta}B\beta\\[0.2em]
   +(A+\frac{B}{2}[e^{i\delta}-B])\beta^2\Big)
   \end{array}
 \\ [2em]
 -\frac{e^{i\delta}}{2\sqrt2}\Big(1+\sqrt3\beta+[A-\frac{1}{2}]\beta^2\Big) ,&
\begin{array}{c}
 \frac{e^{i\delta}}{2\sqrt2}\Big(\sqrt3-[1+2Be^{-i\delta}]\beta\\[0.2em]
 +\sqrt3[A-\frac{1}{2}-\frac{1}{2}B^2]\beta^2\Big)
 \end{array},
 &
 \begin{array}{c}
 \frac{1}{\sqrt2}\Big(1+\frac{\sqrt3}{2}Be^{i \delta}\beta\\[0.2em]
 -[A+\frac{B^2}{2}+\frac{B}{2}e^{i \delta}]\beta^2\Big)
 \end{array}
  \end{array}\right)}
\end{widetext}
Assuming the BF values, $\beta$ and the relation between $A$
and $\delta$ are the same as the results presented in Sec.
\ref{sec:betaSquared}. If $\sin \theta_{13}=0.1$ is used, we obtain
 \dis{B=2.65.}

Without the CP violation, deviation of $\theta_2$ from $\frac{\pi}{4}$ might be of order $\beta$. In this case, $A\simeq 2.2$. Our parametrization of the (13) element being  of $\Od(\beta)$  is given by
  \begin{widetext}

\dis{ \left(
\begin{array}{ccc}
  \frac{1}{2}(\sqrt3-\beta-\frac{\sqrt3}{2}\beta^2),& \frac{1}{2}\Big(1+\sqrt3\beta-\frac{1}{2}(1+B^2)\beta^2\Big),
& \frac{B}{2} (1+\sqrt3 \beta) \beta ,\\[1 em]
\begin{array}{c}
  -\frac{1}{2\sqrt2}\Big(1+(\sqrt3-A)\beta\\[0.2em]
  -[\frac{1}{2}+\frac{A^2}{2}+\sqrt3 A]\beta^2\Big)
   \end{array},&
  \begin{array}{c}
  \frac{1}{2\sqrt2}\Big(\sqrt3-[1+\sqrt3 A-2Be^{-i\delta}]\beta\\[0.2em]
  -(\frac{\sqrt3}{2}(1+A^2+B^2)-A(1+2Be^{-i\delta}))\beta^2\Big)
  \end{array}, &
  \begin{array}{c}
   -\frac{e^{-i\delta}}{\sqrt2}\Big(1-\frac{1}{2}(e^{i \delta}\sqrt3B-2A)\beta\\[0.2em]
   -\frac{1}{2}(A^2+B^2-e^{i\delta}(\sqrt3A+1)B)\beta^2\Big)
   \end{array}
 \\ [2em]
\begin{array}{c}
 -\frac{e^{i\delta}}{2\sqrt2}\Big(1+(\sqrt3+A)\beta\\[0.2em]
 +[\sqrt3A-\frac{A^2}{2} -\frac{1}{2}]\beta^2\Big)
 \end{array},&
 \begin{array}{c}
 \frac{e^{i\delta}}{2\sqrt2}\Big(\sqrt3-[1-\sqrt3A+2Be^{-i\delta}]\beta\\[0.2em]
 -\frac{1}{2}[\sqrt3(1+A^2+B^2)+2A(1-2Be^{-i\delta})]\beta^2\Big)
 \end{array},&
 \begin{array}{c}
 \frac{1}{\sqrt2}\Big(1+(\frac{\sqrt3}{2}Be^{i \delta}-A)\beta\\[0.2em]
 -\frac{1}{2}[A^2+B^2+e^{i \delta}B(1-\sqrt3A)]\beta^2\Big)
 \end{array}
  \end{array}\right)}

\end{widetext}

On the other hand, with the traditional Chau-Keung-Maiani parametrization, we obtain a parametrization where the mixing of $V_{23}$ and $V_{33}$ is maximal.
kept to $\Od(\beta)$, with the (13) element being of order $\beta$, is
\begin{widetext}
\dis{ \left(
\begin{array}{ccc}
  \frac{1}{2}(\sqrt3-\beta-\frac{\sqrt3}{2}(1+B^2)\beta^2),& \frac{1}{2}(1+\sqrt3\beta-\frac{1}{2}(1+B^2)\beta^2),
&B \beta  ,\\[1 em]
   \begin{array}{c}
 -\frac{1}{2\sqrt2}(1+\sqrt3(1+Be^{-i\delta})\beta\\[0.2em]
 -(A+\frac{1}{2}+Be^{-i\delta})\beta^2)
   \end{array},&
    \begin{array}{c}
  \frac{1}{2\sqrt2}\Big(\sqrt3-(1+Be^{-i\delta})\beta\\[0.2em]
  -\sqrt3(A+\frac{1}{2}+Be^{-i\delta})\beta^2\Big) \end{array},
  &\frac{1}{\sqrt2}(1+(A-\frac{B^2}{2})\beta^2)e^{-i \delta}
 \\ [2em]
  \begin{array}{c}
 \frac{1}{2\sqrt2} \Big(e^{i \delta}+\sqrt3(e^{i \delta}-B)\beta\\[0.2em]
 +([A-\frac{1}{2}]e^{i \delta}+B)\beta^2\Big)
 \end{array} ,&
  \begin{array}{c}
  -\frac{1}{2\sqrt2} \Big(\sqrt3 e^{i \delta}-[e^{i \delta}-B]\beta\\[0.2em]
  +\sqrt3([A-\frac{1}{2}]e^{i \delta}+B)\beta^2 \Big)
  \end{array}    ,
  & \frac{1}{\sqrt2}\Big(1-(A+\frac{B^2}{2})\beta^2\Big)
  \end{array}\right)\label{eq:betaform}
  }

\end{widetext}

 This might be useful in fitting the current experimental data. As can be seen in Sec. \ref{sec:exp}, the matrix elements experimentally accessible are $V_{12}, V_{13}$ and $V_{23}$, which can be used to determine $\theta_{{\rm sol}}$ and $\theta_{{\rm atm}}$. Especially, $V_{12}$ and $V_{23}$ are given by $s_{12}c_{13}$ and $s_{23}c_{13}$, respectively, in the Chau-Keung-Maiani parametrization. Since $\cos \theta_{13}$ is very close to unity, and insensitive to the deviation of $\theta_{13}$ from zero as long as $\theta_{13}$ stays very small, measurements of these values give the almost direct information on $\theta_{12}$ and $\theta_{23}$. Moreover, deviation of  $\theta_{23}$ from $\frac{\pi}{4}$ shows  directly the deviation from the bi-maximality. Comparing (\ref{eq:betaform}) with the BF values,
and assuming $\sin \theta_{13}=0.1$, we obtain
\dis{
\beta=0.060,~~B=1.66,~~A=0~~~
}
and the CP phase $\delta$ is undetermined.

 From the complete form of Chau-Keung-Maiani type parametrization, (\ref{eq:CKM}), each term in the determinant is given by

\dis{
V_{11}V_{22}V_{33}& =c_{12}^2c_{23}^2c_{13}^2-c_{12}s_{12}c_{23}s_{23}c_{13}^2s_{13}e^{i \delta}\\
-V_{11}V_{23}V_{32} &=c_{12}^2s_{23}^2c_{13}^2+c_{12}s_{12}c_{23}s_{23}c_{13}^2s_{13}e^{i \delta}\\
V_{12}V_{23}V_{31} &=s_{12}^2s_{23}^2c_{13}^2-c_{12}s_{12}c_{23}s_{23}c_{13}^2s_{13}e^{i \delta} \nonumber}
\dis{-V_{12}V_{21}V_{33} &=s_{12}^2c_{23}^2c_{13}^2+c_{12}s_{12}c_{23}s_{23}c_{13}^2s_{13}e^{i \delta}\\
V_{13}V_{21}V_{32} &=c_{12}s_{12}c_{23}s_{23}s_{13}\cos\delta-c_{12}s_{12}c_{23}s_{23}c_{13}c_{13}^2s_{13}e^{i \delta}\\
&+s_{12}^2c_{23}^2s_{13}^2+c_{12}^2s_{23}^2s_{13}^2
\\
-V_{13}V_{22}V_{31} &-c_{12}s_{12}c_{23}s_{23}s_{13}\cos\delta+c_{12}s_{12}c_{23}s_{23}c_{13}c_{13}^2s_{13}e^{i \delta}\\
&+c_{12}^2c_{23}^2s_{13}^2+s_{12}^2s_{23}^2s_{13}^2.}

In particular with our parametrization,

\begin{widetext}

\dis{
V_{11}V_{22}V_{33} &=\frac{1}{8}\Big((3-2\sqrt3\beta-(2+6A+3B^2)\beta^2)-(\sqrt3+2\beta)B\beta \cos\delta - i(\sqrt3+2\beta)B\beta \sin\delta\Big)\\
-V_{11}V_{23}V_{32} &=\frac{1}{8}\Big((3-2\sqrt3\beta-(2-6A+3B^2)\beta^2)+(\sqrt3+2\beta)B\beta \cos\delta +i(\sqrt3+2\beta)B\beta \sin\delta\Big)\\
V_{12}V_{23}V_{31} &=\frac{1}{8}\Big((1+2\sqrt3\beta+(2+2A-B^2)\beta^2)-(\sqrt3+2\beta)B\beta \cos\delta -i(\sqrt3+2\beta)B\beta \sin\delta \Big)\\
-V_{12}V_{21}V_{33} &=\frac{1}{8}[(1+2\sqrt3\beta+(2-2A-B^2)\beta^2)+(\sqrt3+2\beta)B\beta \cos\delta +i(\sqrt3+2\beta)B\beta \sin\delta]\\
V_{13}V_{21}V_{32} &=\frac{1}{8}\Big(4B^2\beta^2-i(\sqrt3+2\beta)B\beta \sin\delta\Big)
\\
-V_{13}V_{22}V_{31} &=\frac{1}{8}\Big(4B^2\beta^2+i(\sqrt3+2\beta)B\beta \sin\delta\Big)
}

\end{widetext}




\section{Conclusion}
In conclusion, we parametrized the PMNS matrix in terms of small parameter $\beta$ by expanding it around a dodeca symmetry point. Since the (13) element is not measured correctly, we attempt two cases: $\Od(\beta)$ and $\Od(\beta^2)$. The magnitude of (13) element is crucial in understanding the leptonic CP violation.

\acknowledgments{This work is supported in part by the National Research Foundation  (NRF) grant funded by the Korean Government (MEST) (No. 2005-0093841).}

\section*{Appendix A: The CKM matrix parametrization based on dodeca-symmetry\label{sec:CKM}}

To parametrize the PMNS matrix, we expand solar angle in terms of a small parameter $\beta$ around $\frac{\pi}{6}$, predicted by non-Abelian $D_{12}$ symmetry \cite{KimSeo10}. For consistency with this parametrization, it is reasonable to expand the CKM matrix also around the value obtained by this symmetry.
For the specific vacuum considered in Ref. \cite{KimSeo10}, $\theta_{{\rm sol}}=\frac{\pi}{6}$ and  $\theta_{{\rm C}}=\frac{\pi}{12}$ were given. Even though the expansion of the CKM matrix  around the identity is widely used, \ie the trivial vacuum is chosen from the $D_{12}$ point of view, it is also useful to parameterize the CKM matrix around $\frac{\pi}{12}$. Then, the correction must be small, not like $\lambda\simeq 0.23$ \cite{Wol83}. We present such a parametrization below.
\begin{widetext}
\dis{ \left(
\begin{array}{ccc}
  \frac{1+\sqrt3}{2\sqrt2}(1-\frac{\gamma^2}{2})-\frac{-1+\sqrt3}{2\sqrt2}\gamma,& \frac{-1+\sqrt3}{2\sqrt2}(1-(1+\kappa_b^2)\frac{\gamma^2}{2})+\frac{1+\sqrt3}{2\sqrt2}\gamma,
&\kappa_b\gamma( \frac{-1+\sqrt3}{2\sqrt2}+ \frac{1+\sqrt3}{2\sqrt2}\gamma)  ,\\[1 em]
   -(\frac{-1+\sqrt3}{2\sqrt2}(1-(1+\kappa_t^2)\frac{\gamma^2}{2})+\frac{1+\sqrt3}{2\sqrt2}\gamma),&
\begin{array}{c}
\frac{1+\sqrt3}{2\sqrt2}\Big(1-(1+\kappa_b^2+\kappa_t^2)\frac{\gamma^2}{2}\Big)\\
-\frac{-1+\sqrt3}{2\sqrt2}\gamma
+e^{-i\delta}\kappa_t\kappa_b\gamma^2
\end{array},
  & \kappa_b\gamma(\frac{1+\sqrt3}{2\sqrt2}-\frac{-1+\sqrt3}{2\sqrt2}\gamma)-e^{-i\delta}\kappa_t \gamma
 \\ [2em]
  -e^{i\delta}\kappa_t \gamma(\frac{-1+\sqrt3}{2\sqrt2}+\frac{1+\sqrt3}{2\sqrt2}\gamma) ,&
  e^{i\delta}\kappa_t\gamma(\frac{1+\sqrt3}{2\sqrt2}-\frac{-1+\sqrt3}{2\sqrt2}\gamma)-\kappa_b\gamma ,
  & 1-(\frac{1}{2}(\kappa_b^2+\kappa_t^2)-e^{i\delta}\kappa_t\kappa_b\frac{1+\sqrt3}{2\sqrt2})\gamma^2
  \end{array}\right)\nonumber
  }

\end{widetext}
where $\gamma$ is estimated to be $-0.034, |\kappa_b|=0.59$, and $ |\kappa_t|=1.08$.

 \section*{Appendix B: Majorana  phases and neutrinoless double beta decay \label{sec:maj}}

  If Majorana phases are taken into account, PMNS matrix should be modified by multiplying
     \dis{ \left(
\begin{array}{ccc}
  1& 0&0\\[1 em]
  0& e^{i \Delta \alpha_{21}}  &0\\ [1 em]
0& 0 & e^{i \Delta \alpha_{31}}
  \end{array}\right)\label{eq:CKM}
  }
on the right side. Such phases can be measured through neutrinoless double beta ($0\nu \beta \beta$)decay \cite{0vbb}, $(Z,A)\to(Z \pm 2, A)+2 e^{\mp}$. The $0\nu\beta\beta$ decay rate is proportional to the squared effective neutrino mass,
\dis{\langle m_{\beta \beta} \rangle ^2= \vert \sum_i V^2_{ei} m_{\nu_i} \vert^2}
and in terms of exact form of PMNS matrix element, it is given by
\dis{\vert c_1^2 m_1^2 + s_1^2 c_3^2 e^{i \Delta \alpha_{21}} m_2^2+s_1^2s_3^2 e^{i \Delta \alpha_{31}}m_3^2 \vert^2}
in the parametrization (\ref{eq:KSform}) and
\dis{\vert c_{12}^2 c_{13}^2 m_1^2 + s_{12}^2 c_{13}^2 e^{i \Delta \alpha_{21}} m_2^2+s_{13}^2 e^{i (\Delta \alpha_{31}-\delta)}m_3^2 \vert^2}
in the Chau-Keung-Maiani parametrization, (\ref{eq:CKM}).

\vskip 0.5cm

\end{document}